\documentstyle{mn}
 at 10truept
\title
     [Design and analysis of two-dimensional redshift surveys]
{\vglue-3.0truecm
\vglue 2.5truecm
            Design and analysis of redshift surveys
\author
     [A.F. Heavens and A.N. Taylor]
     {A.F. Heavens and A.N. Taylor\\
     Institute for Astronomy, 
     University of Edinburgh,
     Royal Observatory,
     Blackford Hill, 
     Edinburgh, 
     U.K.}}
\def\bib{\parskip=0pt\par\noindent\hangindent\parindent
    \parskip =2ex plus .5ex minus .1ex}
\newcommand{\be}{\begin{equation}}
\newcommand{\ee}{\end{equation}}
\newcommand{\ba}{\begin{eqnarray}}
\newcommand{\ea}{\end{eqnarray}}
\newcommand{\gs}{\mathrel{\raise1.16pt\hbox{$>$}\kern-7.0pt 
\lower3.06pt\hbox{{$\scriptstyle \sim$}}}}         
\newcommand{\ls}{\mathrel{\raise1.16pt\hbox{$<$}\kern-7.0pt 
\lower3.06pt\hbox{{$\scriptstyle \sim$}}}}         
\newcommand{\kb}{{\bf k}}
\newcommand{\rb}{{\bf r}}
\newcommand{\nn}{\nonumber \\}
\newcommand{\vlg}{{\bf v}_0}
\newcommand{\bfv}{{\bf v}}
\renewcommand{\sb}{{\bf s}}
\newcommand{\mn}{{MNRAS}}

\include{epsf}

\begin{document}

\maketitle

\begin{abstract}
In this paper we consider methods of analysis and optimal design of
redshift surveys.  In the first part, we develop a formalism 
for analysing galaxy redshift surveys which are essentially 
two-dimensional, such as thin declination slices.  
The formalism is a power spectrum method, using spherical coordinates,
allowing the distorting effects of galaxy peculiar velocities to be 
calculated to linear order on the assumption of statistical isotropy
but without further approximation.  
In this paper, we calculate the measured two-dimensional power for 
a constant declination strip, widely used in redshift surveys.  
We present a likelihood method for estimating the three-dimensional 
real-space power spectrum and the redshift distortion simultaneously,
and show that for thin surveys of reasonable depth, the large-scale
3D power cannot be measured with high accuracy.   The redshift distortion
may be estimated successfully, and with higher accuracy if the 3D power 
spectrum can be measured independently,  for example from a large-scale 
sky-projected catalogue.

In the second part, we show how a 3D survey design can be optimized to measure
the power spectrum, considering whether areal coverage is more important 
than depth, and whether
the survey should be sampled sparsely or not.  We show quite
generally that width is better than depth, and show how the optimal
sparse-sampling fraction $f$ depends on the power $P$ to be measured.
For a Schechter luminosity function, a simple optimization 
$fP \simeq 500 h^{-3}$ Mpc$^3$ is found.

\end{abstract}

\section{Introduction}

The measurement of fluctuation power in the galaxy distribution is an
important test of galaxy formation models, since the fluctuation
spectrum, of mass at least, is predicted readily by such models.
Power can be measured from three-dimensional redshift surveys, or from
projected catalogues, by numerical inversion techniques.    Redshift
surveys have the advantage (and disadvantage) that they are distorted
by the effects of peculiar velocities, and can therefore be used to
extract information on the density parameter, under the assumption that
structure grows by gravitational instability.     One has then the
possibility of measuring three-dimensional power and the density
parameter (via $\beta \equiv \Omega_0^{0.6}/b$, where $b$ is the bias
parameter for the survey in hand) from a galaxy redshift survey
(Heavens \& Taylor 1995; hereafter HT95;  see also Kaiser 1987, 
Hamilton 1992, Cole et al. 1994).

It is clear that the longest wavelength which can be measured is
limited by the size of the survey, so it is attractive to consider
surveys which are essentially one- or two-dimensional,  to maximise at
least one dimension without incurring prohibitive cost in observation
time (e.g. Broadhurst et al 1990).  The difficulty with such an
approach as a method for measuring the power spectrum is that a
low-dimensional power measurement at a given wavenumber will have a
contribution (which may be dominant) from much smaller scales in three
dimensions (e.g. Kaiser \& Peacock 1991).  The interpretation of the
observed power spectrum can therefore be difficult.  For surveys which do
not correspond to the `distant-observer' approximation (cf. Kaiser 1987),
the power spectrum measurement and the redshift distortion become linked,
and this further complicates the analysis.  

The ease with which the parameters of interest may be extracted
depends on the choice of coordinate system and basis functions in which
the density field is expanded.  
For two reasons the choice of spherical polar coordinates is
compelling.  Firstly the survey is almost certain to be defined in
terms of a fixed areal coverage (independent of depth), and secondly,
a flux limited survey will have a selection function $\phi$ 
(or, equivalently, a mean observed density $\rho_0(r)$) which is
dependent on distance, but not on direction.  The mean density of the
survey is then separable in spherical coordinates $\bar \rho({\bf r}) =
\rho_0(r)M(\theta,\varphi)$, where $M$ is either 1 or 0 depending on
whether the direction ($\theta,\varphi$) is in the survey or not.
The second reason is that, unless $\beta \ll 1$, it is impossible to
ignore redshift distortion effects, and since the distortion between
the real-space map and the redshift-space map is purely radial, it is
straightforward to include the distortion in a power-spectrum analysis
in spherical coordinates (cf Zaroubi \& Hoffman 1996).

The choice of basis functions must also be done with some care.  It is
very useful to choose functions which pick up a narrow range of
wavenumbers from three-dimensional space.  In this regard, spherical
Bessel functions are ideal, and they also have advantages in that the
redshift distortion is relatively simple in this system.   An arbitrary
choice of functions (or equivalently, an arbitrary choice of projection
onto the sky) makes the interpretation of measured 2D power difficult,
since the 3D wavenumbers contributing to the power may be poorly
constrained.    Even in this idealized case, the range of 3D power
contributing to the 2D coefficients can be large, especially for thin surveys
of a few degrees thickness.  In these cases, 3D power spectrum at large 
scales becomes difficult to achieve with high accuracy.

One of the aims of the paper up to this point is to demonstrate that
the expected accuracy of a proposed survey in measuring some parameter 
can be estimated in advance, and this sort of study can and should influence
how a survey is designed, whether in 2D or 3D.
We investigate in section 4 how 3D design may be optimised for power 
spectrum estimation,  subject to constraints.   We find that, if observing
time is constrained, then it is always better to cover a large area
on the sky, rather than going deep.  In some cases it can be advantageous to
sample galaxies sparsely,  with the sparse-sample fraction being dependent
on the power to be measured and the luminosity function of the galaxies.
We present a very simple formula for calculating the optimal  sparse-sample
fraction.   

The paper is laid out as follows:  in Section 
2 we consider power measurements in idealised cases of 1D 
and 2D surveys, which highlight the problems which such surveys 
have to address.  
In Section 3 we present a new method for analysing constant declination strips,
for measuring the real-space power spectrum and redshift distortion.  
In Section 5, we solve the problem of 
designing surveys optimised for 3D power estimation.

\section{1D and 2D: general considerations}

The fractional overdensity is $\delta(\rb) \equiv \rho(\rb)/\bar\rho -1$, 
where $\bar\rho$ is the mean density.  Our Fourier transform convention is 
$\delta_\kb \equiv  \int d^3\rb \,\delta(\rb)\,e^{-i\kb.\rb}$
with inverse $\delta(\rb) \equiv {1\over (2\pi)^3} \int \,d^3\kb
\,\delta_\kb\,e^{i\kb.\rb}$.  The Power spectrum is defined by
\be
\langle\delta_\kb\delta_{\kb'}^*\rangle \equiv 
(2\pi)^3 P_{3D}(k)\,\delta^D(\kb-\kb')
\label{power}
\ee
so that the correlation function is
\be
\xi(r) = {1\over (2\pi)^3}
\int\,d^3\kb\,P_{3D}(k)\,e^{i\kb.\rb}.
\ee
In an idealized pencil-beam survey, the density field along a line is measured,
and the 1D power spectrum estimated.  To relate this to the 3D power spectrum,
we follow Lumsden, Heavens \& Peacock (1987), noting that the correlation 
function is the same along the line as in 3D, by isotropy.
\ba
P_{1D}(\tilde k)\! & = & \int\,dx\,\xi(x)\exp(-i\tilde k x) \nn
 \!\!\!\!\!\! & = & \! {1\over (2\pi)^3}\int\,dx d^3\kb\,
             P_{3D}(k)\exp(ik_xx)\exp(-i\tilde k x)\nn
\ea
where we assume the pencil beam lies along $y=z=0$.   The integration over $x$ 
gives a 1D delta function, and changing the remaining integration over 
$k_x$ and $k_y$ to a polar integration, we get
\be
P_{1D}(k) = {1\over (2\pi)^2}\,\int_{|k|}^\infty\,d\bar{k}\,P_{3D}(\bar{k})
\, \bar{k}
\ee
Hence we see that power in one dimension 
comes from {\it all} shorter wavelengths in 3D.  This is readily 
understandable by consideration of the following 2D $\rightarrow$ 1D 
illustration:   imagine looking along a corrugated roof 
at an angle to the corrugations.  The separation 
of peaks along the line is longer in 1D by a geometrical factor,
so the 1D power (averaged over angles) has contributions from 2D power
at all shorter wavelengths.
  
Notice also that the 1D power spectrum must be a monotonically 
decreasing function of $k$  (the
squared amplitudes of the Fourier coefficients may not be monotonic, 
being drawn from a Rayleigh distribution).
Also note that if the 3D power spectrum 
has a cutoff at some large wavelength, the 1D power spectrum 
will be constant (and non-zero) on all larger scales.

For comparison with later analysis in this paper, we define 
the kernel $G(k,\bar{k})$ such that the measured power is
\be
P(k) \equiv \int_0^\infty\,d\ln \bar{k}\,P_{3D}(\bar{k})\,G(k,\bar{k})
\ee
from which we see that the kernel for a 1D skewer is
\be
G(k,\bar{k}) = 2\pi\,{\bar{k}}^2\,\Theta(\bar{k}-k)
\ee
where $\Theta$ is the Heaviside function.   This unpleasant 
convolution function is shown in Figure 1 (throughout we plot 
$G(k,\bar k)/\bar k$, since we use a linear rather than logarithmic $k$ axis).
For a practical 
pencil-beam survey, the kernel 
will be suppressed at high $k$,  but this calculation 
illustrates the severe problems in interpreting the power 
spectrum of pencil-beam surveys -- the 1D power may be coming 
from much larger wavenumbers in 3D.

%
%
\begin{figure}
\begin{center}
\setlength{\unitlength}{1mm}
\begin{picture}(90,70)
\includegraphics{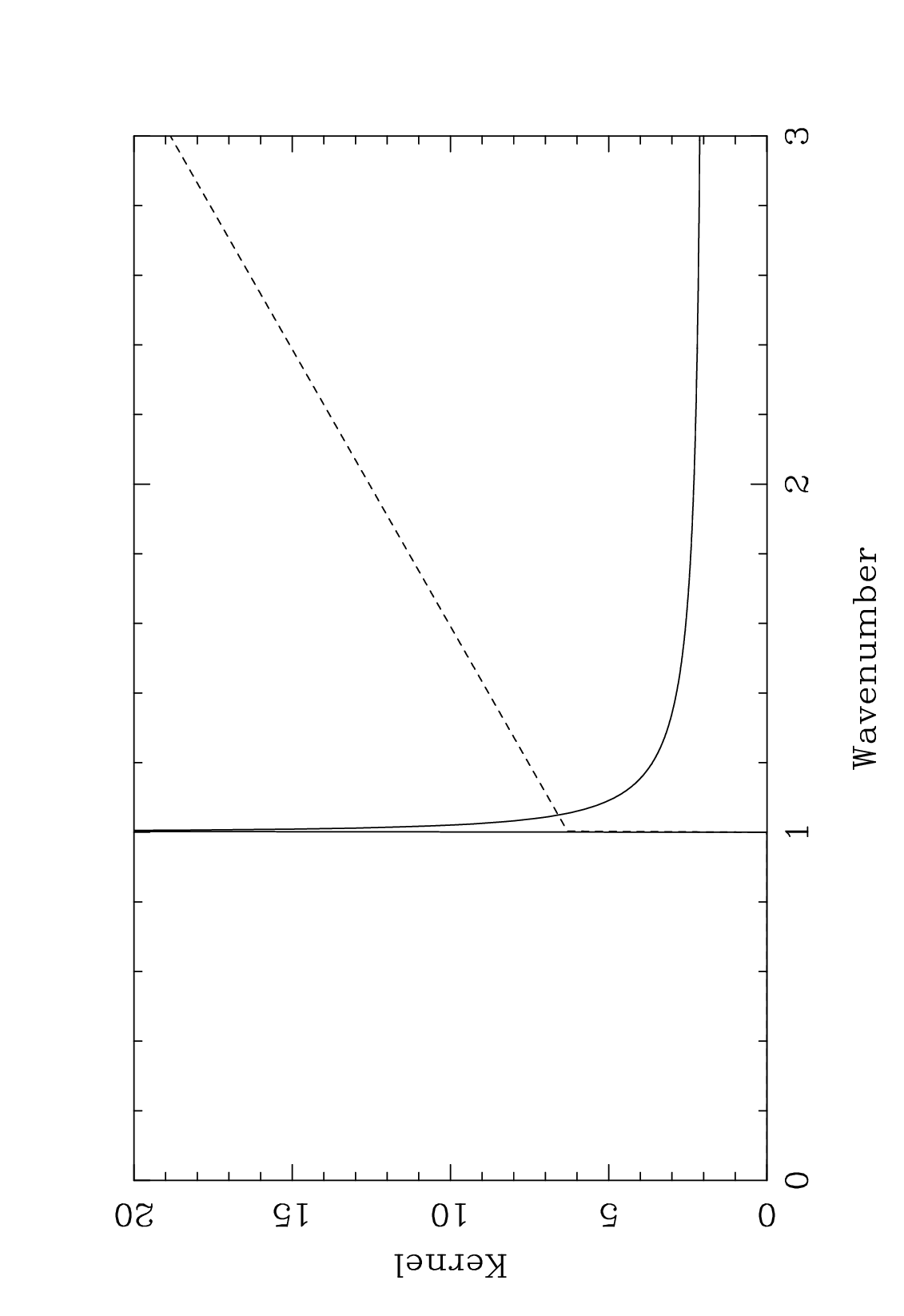} 
\end{picture}
\end{center}
\vspace{2cm}
\caption{Kernel function $G(k,\bar k)/\bar k$ for a thin sheet 
(solid) and pencil-beam survey 
(dashed).   The 2D wavenumber is unity.}
\end{figure}
%
%

For thin, infinite  plane surveys, a similar analysis (Peacock 1991) gives
\be
P_{2D}(k) = 2\int_k^\infty\,d\bar{k}\, P_{3D}(\bar{k})\,{\bar{k}\over
\sqrt{{\bar{k}}^2-k^2}}.
\ee
The associated kernel $G(k,\bar{k}) = 2{\bar{k}}^2 \Theta(\bar{k}-k) /
\sqrt{{\bar{k}}^2-k^2}\,$ is 
also shown in Figure 1.    Again what one finds is that the kernel feeds 
a lot of
power (if it exists) from high $k$ to low $k$.  If the plane survey is of 
finite thickness, the high-$k$ part of the kernel will be suppressed, 
and weighting of the data can help still further (see Section 3) but it is 
still possible that 2D power at $k$ may have little connection with 3D power 
on such scales.

\section{2D surveys}

For surveys which have one dimension considerably smaller 
than the other two, it is sensible to reduce the dimensionality of 
the survey by projection onto a 2D surface.   We can then reduce the 
dimensionality of the transform correspondingly.  As emphasized in the
introduction, there are considerable advantages in using spherical 
coordinates for the analysis.   The survey is almost certainly characterized
by a radial selection function, and an angular selection defining the
boundary (cf Fabbri \& Natale 1990, Scharf et al. 1992, Scharf \& Lahav 1993, 
Fisher et al 1994a,b 1995).  Also, the effects of redshift distortion enter 
only in the
radial direction.   In the case of a thin slice survey, one avoids 
any difficulties which might be apparent in `flattening-out' the survey into
a plane suitable for cartesian analysis.  One final point is that it is the
largest scale modes where the radial nature of the distortion is most 
important to treat correctly.

In this section, we develop a 2D expansion of the density field projected
onto a fixed declination, allowing for redshift distortions.   We use
coordinates $s, \theta, \varphi$, where $s = cz/H_0$ is the 
distance assigned on the basis of the redshift $z$, assuming 
uniform expansion ($H_0$ is the Hubble constant).  This is the
normal distance assigned in redshift surveys, and it differs 
from the true distance $r$ because of peculiar velocities.
In this paper we consider only $z \ll 1$, but the spherical coordinate 
system allow the effects of non-euclidean geometry and temporal evolution
to be included if desired.

Our 2D expansion is based on the 3D Fourier-Bessel expansion of
the density field (cf HT95, Lahav 1993):
\be
\delta_{\ell m}(k) = \sqrt{{2\over \pi}}\,\int\,d^3\rb\,
\delta(\rb)\,j_\ell(kr)\,Y_\ell^{m*}(\theta,\varphi)
\ee
with inverse
\be
\delta(\rb) =  \sqrt{{2\over \pi}}\,\sum_{\ell,m}\int\, dk\,k^2 
\delta_{\ell m}(k)\,j_\ell(kr)\,Y_\ell^m(\theta,\varphi).
\label{delta}
\ee
The statistical properties of $\delta_{\ell m}(k)$ are 
derived in the appendix.  We proceed by projecting all galaxies
onto the central value $\theta=\theta_0$, and expand in terms
of $m$ and $k$.   At this stage, we leave the choice of 
radial expansion function general, $f(kr)$.  We also allow for radial
and angular weighting of the data via the functions $w_s({\bf s})$ and
$w_\Omega(\Omega)$, which may help in optimizing the signal-to-noise
and apodizing.  The radial weight may be $k$-dependent.
For a constant declination strip, the obvious orientation of
coordinates is to have the centre of the strip at constant
$\theta=\theta_0$, and to expand in terms of $m$.  We let the thickness
be $\Delta\theta$, and the width $\Delta\varphi$, centred on
$\varphi=0$.    Our choice of expansion is
\be
\tilde\rho_m(k) \equiv \sqrt{{2\over \pi}}\int\,d^3\sb \rho(\sb)\,f(ks)
\exp(-im\varphi)\,w_s(s)w_\Omega(\Omega).
\ee
As an important aside, there is an issue over which frame of 
reference should be used for 
redshift-space expansions of this sort.  Should the redshift be 
measured in the Local Group frame or the Microwave
Background frame?   In either case, the redshift distance is
\begin{equation}
\sb(\rb) = \rb\left[1+{\left(\bfv-\vlg\right)\cdot \rb \over H_0 r^2}\right]
\end{equation}
where ${\bf v}_0$ is the peculiar velocity of the frame of reference.
This relationship is general, but since we wish to make a perturbation
expansion, we must ensure that the second term in the square brackets 
is always small.  Assuming a sufficiently coherent velocity field
such that {\bf v} approaches the Local Group velocity as $r\rightarrow
0$, we see that the expansion must be done in the Local Group frame.

The difference between the expansion coefficients and their mean values
\be
\rho_m^0(k) = \sqrt{{2\over \pi}}\int d^3\rb \rho_0(r)\,f(kr)
\exp(-im\varphi)\,w_s(r)w_\Omega(\Omega)
\ee
can be related to the $\delta_{\ell m}(k)$ by substituting for
$\delta(\rb)$ from (\ref{delta}), and noting that number conservation
implies that $\rho(\rb)d^3\rb=\rho({\bf s})d^3{\bf s}$:
\ba
D_m(k) & \equiv & \tilde\rho_m(k)-\tilde\rho_m^0(k)\nn
& = & \sum_{\bar\ell\bar m}\,W_{\bar\ell}^{m\bar m}
\int_0^\infty d\bar k\,\delta_{\bar\ell\bar m}(\bar k)\,
\Lambda_{\bar\ell}(k,\bar k)\,\bar k^2
\ea
where
\ba
W_{\bar\ell}^{m\bar m} & \equiv & \sqrt{{(2\bar\ell+1)\over 4\pi}
{(\bar\ell-\bar m)!\over (\bar\ell+\bar m)!}}\,(-1)^{(\bar m +|\bar m|)/2}\times\nn
& & \!\!\!\!\!\!\!\!\!\!\!\!\!\!
{2\sin\left[(\bar m-m)\Delta\varphi/2\right]\over (\bar m - m)}\,
\int_{\cos(\theta_0+\Delta\theta/2)}^{\cos(\theta_0-\Delta\theta/2)}\,d\mu
\,P_{\bar\ell}^{|\bar m|}(\mu)
\ea
and
\ba
\Lambda_{\bar\ell}(k,\bar k) & = & \Phi_{\bar\ell}(k,\bar k)+ 
\beta V_{\bar\ell}(k,\bar k)\nn
\Phi_{\bar\ell}(k,\bar k) & \equiv & \!{2\over\pi}\,\int_0^\infty \,dr
\,\rho_0(r)\,j_{\bar\ell}(\bar kr) f(kr)\,w_s(r)\,r^2\nn
V_{\bar\ell}(k,\bar k)& \equiv & \!{2\over\pi \bar k^2}\int_0^\infty
\! dr \,\rho_0(r)
{d\over dr}\left[f(kr)w_s(r)\right]\times \nn
& & {d\over dr}\left[j_{\bar\ell}(\bar k r)\right] r^2.
\ea
The signal part of the covariance matrix can be written as
\be
\langle D_m(k) D_{m'}^*({k'})\rangle \equiv \int d\ln\bar k
\,P_{3D}(\bar k)\,G_{mm'}(k,{k'},\bar k)
\ee
where
\be
G_{mm'}(k,{k'},\bar k) = (2\pi)^3 \sum_{\bar\ell} Z_{\bar\ell}^{mm'}
 \Lambda_{\bar\ell}(k,\bar k)\Lambda_{\bar\ell}({k'},\bar k) \bar k^3
\ee
and $Z_{\bar\ell}^{mm'}\equiv \sum_{\bar m} W_{\bar\ell}^{m\bar m}
W_{\bar\ell}^{m'\bar m*}$.   The shot noise contribution to the
covariance matrix is 
\ba
\langle D_m(k) D_{m'}^*({k'})\rangle_{\rm SN}&  = & {2\over \pi} \int dr
d\theta d\phi\nn
& &  \!\!\!\!\!\!\!\!\!\!\!\!\!\!\!\!\!\!\!\!\!\!\!\!\!\!\!\!
\!\!\!\!\!\!\!\!\!\!\!\!\!\!\!\!\!\!\!\!\!\!\!\!\!\!\!\!
r^2 \sin\theta \rho_0({\bf r}) f(kr)f(k'r)\exp
\left[i(m-m')\varphi\right] w_s^2(r) w_\Omega^2(\Omega).\nn
\ea
In practice, one splits $D_m(k)$ into real
and imaginary parts, with similar, but more cumbersome, expressions for
the covariance matrix elements.
Some kernels are shown in Fig. 2 and 3  for a survey with a
gaussian selection function $\exp\left[-(r/r*)^2\right]$, with $r*=450 
h^{-1}\,$ Mpc, 
with survey limits $\Delta\theta=6^\circ$ centred at declination 30$^\circ$,
and width $\Delta\varphi=90^\circ$.  
The radial expansion function is chosen to be a spherical 
Bessel function, with $\ell=2$.   This choice is motivated in two ways.  
Firstly, we know that in 3D the Bessel functions give narrow kernels, so
they seem a good start in 2D.  A second, related point, is that the 
function gives little weight to the very nearby part of the survey.
Since this is the thinnest part, it is likely to contribute 
significantly to aliasing difficulties.  The weighting scheme chosen
also helps in this regard.   Fig. 3
shows kernels for plane waves with almost the same wavenumbers as Fig. 2, and 
with direction along the central $\varphi$ value, to correspond as closely 
as possible to Fig. 2 (cf the analysis of the Las Campanas survey by 
Landy et al. 1996).  These curves are simply integrals of the squared
modulus of the window function transform, calculated via a 200$^3$ FFT,
which accounts for their slightly ragged nature.
The comparison between the methods is 
not quite straightforward, as the 2D modes look rather different. Note that
the Fourier modes assume $\beta=0$;  for non-zero $\beta$, the kernels
are extremely complicated in the Fourier case.
The comparison is most stark if one compares the dimensionality of the
objects which one needs to calculate to include redshift distortions.
In essence,  the 2D coefficients are linear combinations of the 3D
coefficients.   If no simplification is possible,
one needs to calculate a 5D object to calculate a range of 2D coefficients.
This is required if one uses Fourier modes  (Zaroubi \& Hoffman 1996,
equation 10), but using the spherical modes
reduces the dimensionality such that the most complicated objects are only 3D
(see (13)).   It is this fact that the kernels for 
non-zero $\beta$ can be readily calculated 
for the spherical modes which is their major advantage.  It arises, 
of course, from the radial nature of the distortion and selection
function, and the use of angular coordinates to delimit the survey.

For high $m$ the spherical kernels are sometimes not centred on $k$, and the 
2D power may come principally from shorter wavelengths in 3D.  
The effects of this, and shot noise and cosmic variance  
can be accounted for correctly using likelihood
techniques,  so the 3D power spectrum can be estimated, but it is clear from
Figs. 2 and 3 that the task is not going to be easy, whichever method is used.
The accuracy with which the power and $\beta$ determination can be done 
with the Fourier-Bessel transform is explored in the next sections. 

%
%
\begin{figure}
\begin{center}
\setlength{\unitlength}{1mm}
\begin{picture}(90,70)
\includegraphics{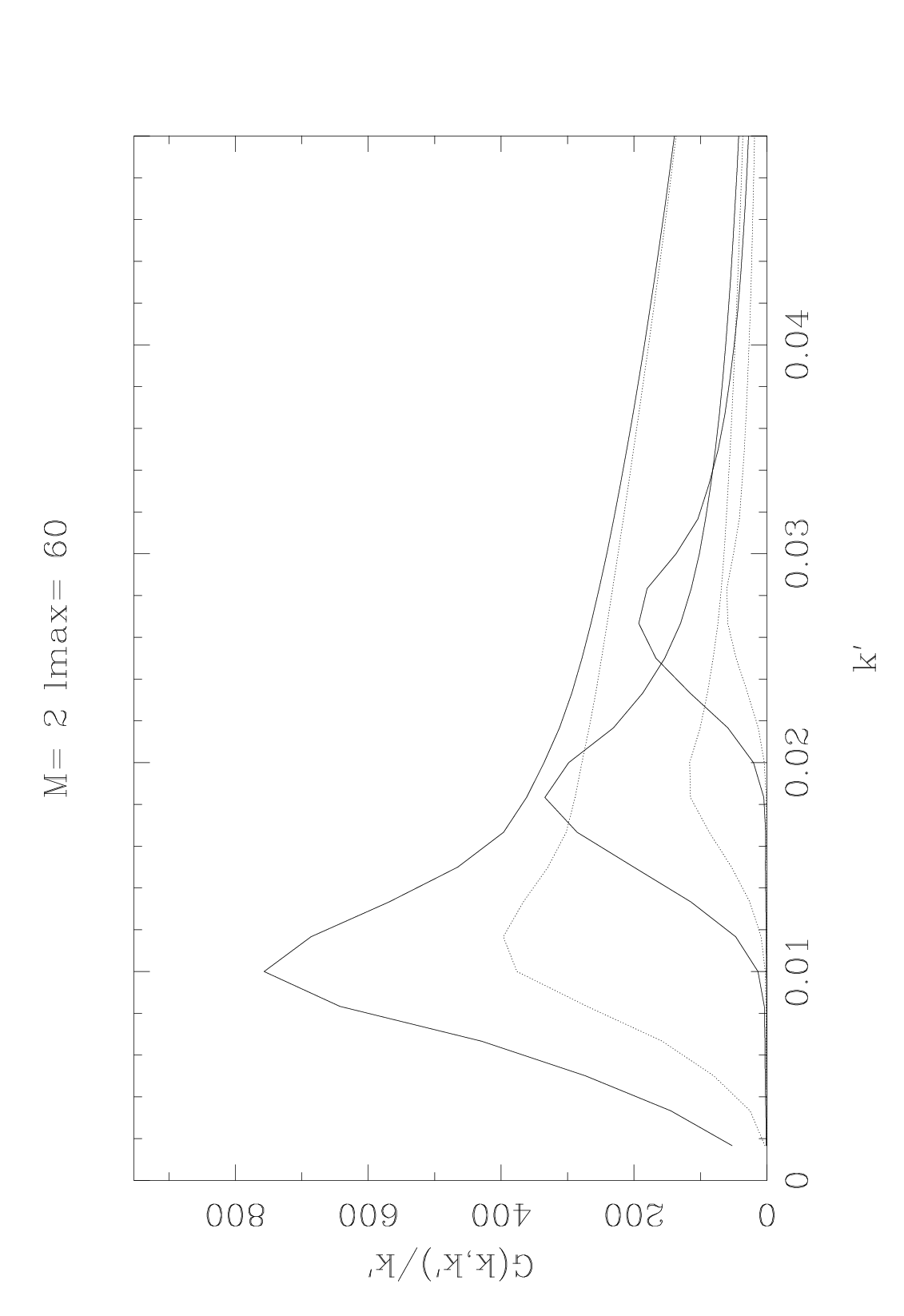}
\end{picture}
\end{center}
\vspace{2cm}
\caption{Kernel function for constant declination slice, for modes
with $m=2$, and, from left to right, $k=0.008,\ 0.016,\ 0.025 h$ Mpc$^{-1}$.  
Solid lines are for
$\beta=1$, dotted for $\beta=0$.}
\end{figure}
%
%

%
%
\begin{figure}
\begin{center}
\setlength{\unitlength}{1mm}
\begin{picture}(90,70)
\includegraphics{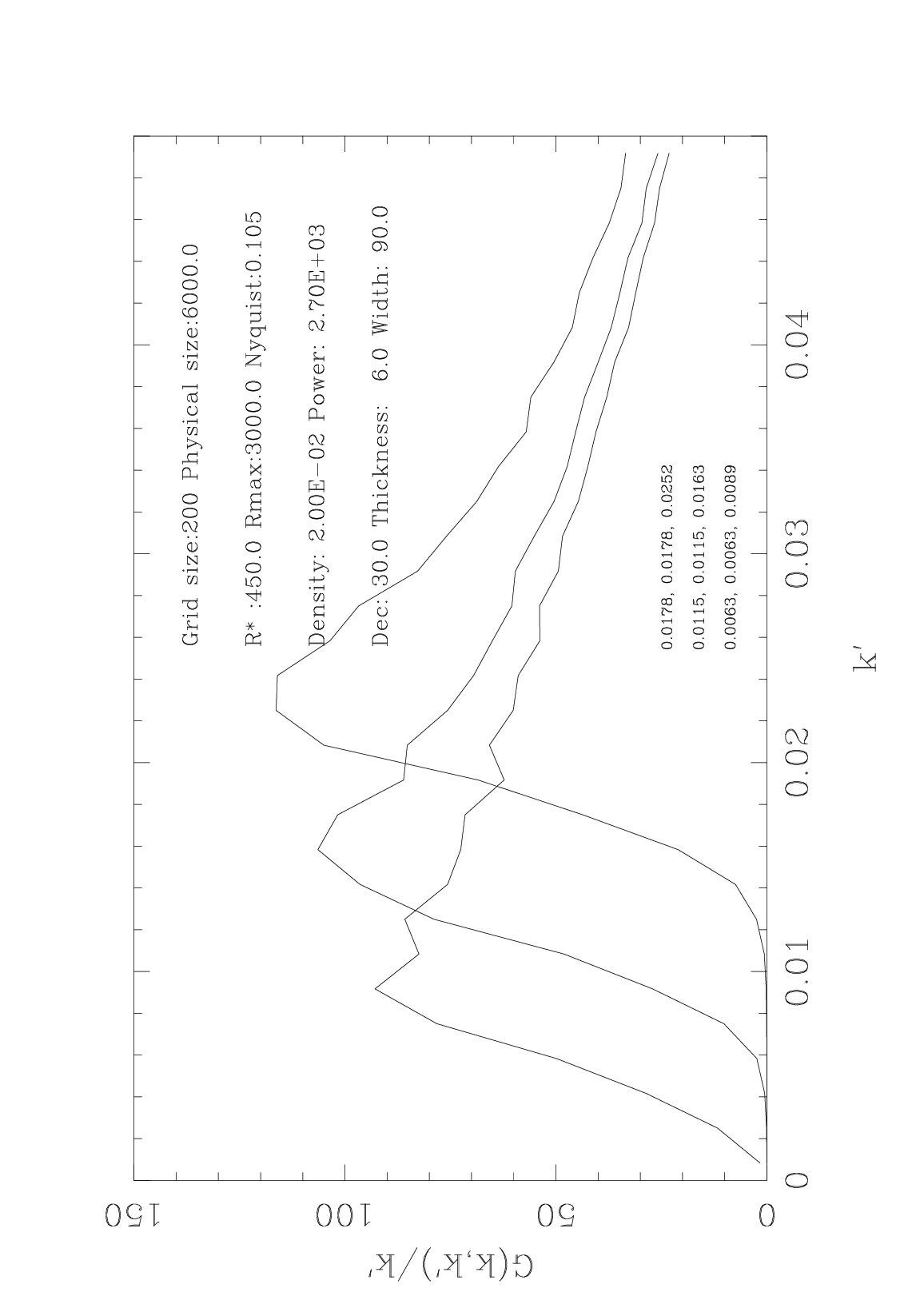}
\end{picture}
\end{center}
\vspace{2cm}
\caption{Kernel function for Fourier modes with same wavenumbers as Fig. 2.}
\end{figure}
%
%
  
\subsection{Parameter estimation}

We can use the analysis method presented in the last section to 
estimate the real space power spectrum and $\beta$
maximizing the likelihood.  Symbolically
\be
{\cal L}(\beta,P) = {1\over (2\pi)^{N/2} ||C||^{1/2}}\exp
\left(-{1\over 2}\sum_{\mu\nu} D_\mu C^{-1}_{\mu\nu} D_\nu\right).
\ee
where $C$ is the covariance matrix of the $N$ data values, dependent on
the (parametrized) $P(k)$ and $\beta$.  Once again,  the data are
the real and imaginary parts of $D_m(k)$.   The likelihood method has the
advantage that all the aliasing effects are treated correctly, and we do
recover 3D power estimates with correct error bars.
   We illustrate this method by analysing a numerical simulation
created with Couchman's AP$^3$M code (Couchman 1991).  A slice between
declinations 20 and 40 degrees is projected onto declination 30, with
a right ascension range of 90 degrees.  The power spectrum is a power-law
$P(k) \propto k^{-1}$, and the likelihood for the amplitude of $P(k)$ and 
$\beta$ is shown in Figure 4.   The details of the analysis are that the nonlinear 
wavenumber (where $k^3 P(k)/(2\pi^2)=1$) is 183 (units are arbitrary), and
the analysis examines modes up to $k_{\rm max}=30$.  
Pushing the maximum analysed 
wavenumber beyond this pushes $\beta$ down, as the effects of Fingers-of-God
become apparent.  These could be reduced by including the effects of
smoothing (cf. HT95), but they have not been 
incorporated here.   The $m$ modes are analysed in steps of 2 from 2 to 20,
and the wavenumbers selected are from 6 to 30 in steps of 2.  There is no
difficulty in principle in taking every $m$ and $n$ mode, but there is a 
numerical problem as adjacent modes are
too strongly correlated and the matrix becomes numerically singular.   
Note that modes separated by 2 are correlated, and the correlations 
correctly accounted for in (19).
We also put a constraint on the wavenumber
perpendicular to the line-of sight, ensuring that it is not too nonlinear,
by rejecting modes with $m\,k/4.0 < k_{\rm max}$.  This ensures that the
transverse wavenumber at the peak of $j_2(kr)$ is no more than 4.0/3.3 
times $k_{\rm max}$.   Experimentation shows that this gives unbiased
estimation of $\beta$ and $P(k)$.   The galaxies are weighted with the Feldman 
et al. (1994) optimized weighting $w_s(r) = 
\left[1+\rho_0(r)P(k)\right]^{-1}$
and $P(k)$ in the weighting is taken as a constant, comparable to the
true power in the simulation.   
The true parameters are shown by the encircled cross.
We see that the method is capable of determining $\beta$ and the power
spectrum with somewhat larger errors than a fully 3D survey (HT95) 
with similar numbers of objects, and also note that here
we need to examine larger wavelengths than in the 3D case (up to the 
nonlinear wavenumber/6, as opposed to nonlinear wavenumber/3 in the 
3D case).  Failure to do this leads to underestimation of $\beta$
because of nonlinear effects.

%
%
\begin{figure}
\begin{center}
\setlength{\unitlength}{1mm}
\begin{picture}(90,70)
\includegraphics{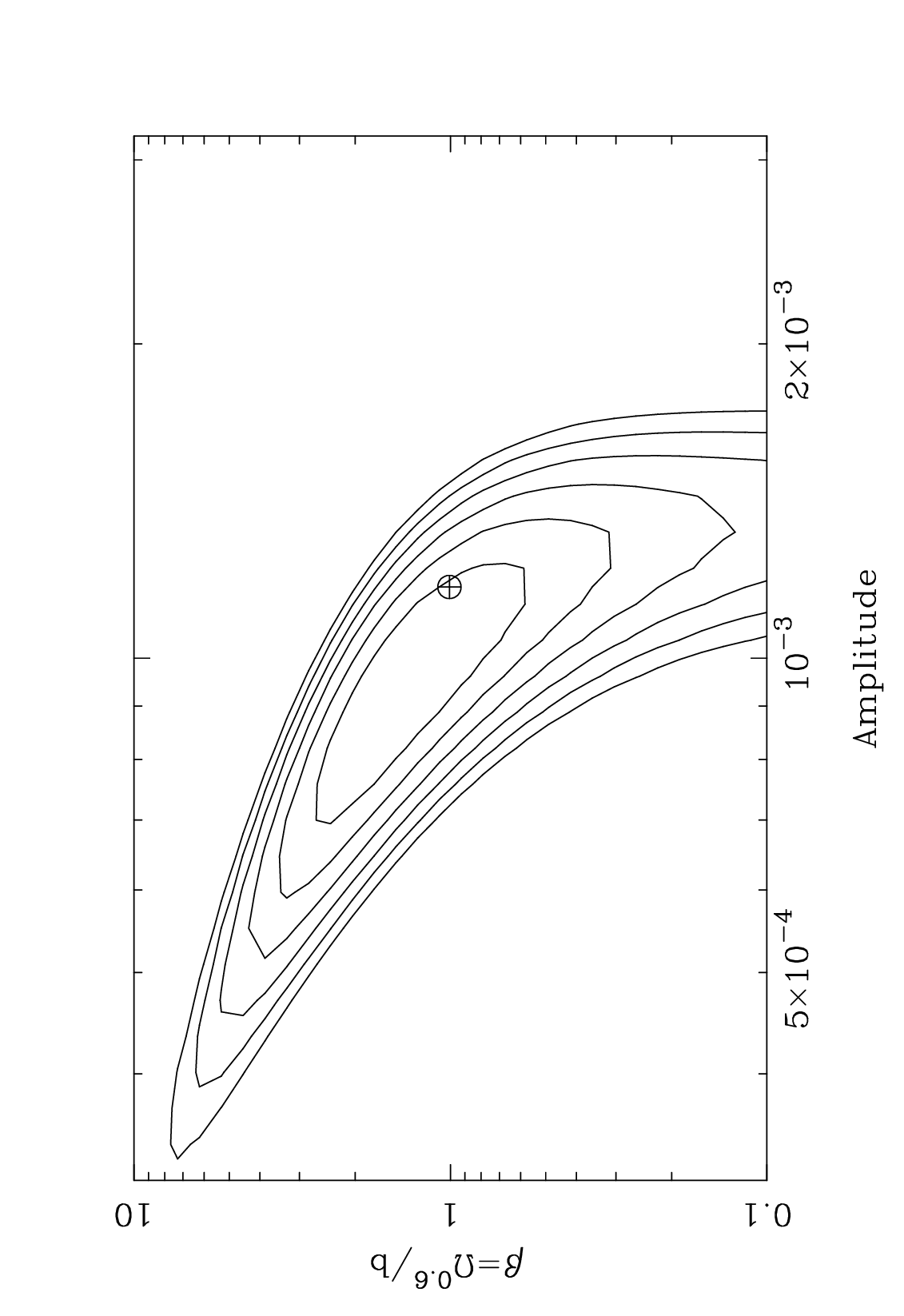} 
\end{picture}
\end{center}
\vspace{2cm}
\caption{Likelihood function for the amplitude of the real-space power
spectrum and $\beta$ for a numerical simulation whose true parameters
are shown by the cross.  The contours are separated by 0.5 in ln(likelihood).}
\end{figure}  
%
%

\subsection{Errors on $\beta$ and $P(k)$}

In this section, we calculate the expected errors in a deep, 
thin-slice survey, such as might be achievable in the first year of 
the AAT 2dF survey.    The error on the parameters is readily estimated using
the Fisher information matrix (Tegmark, Taylor \& Heavens 1996). 

For a set of parameters $\theta_i,\ i=1,N$ (e.g. $\beta$ and the
power spectrum in $N-1$ wavenumber bins), the covariance matrix of the
parameter estimates is
\be
{\bf T} = \langle {\bf \theta \theta}^T \rangle -
\langle {\bf \theta}\rangle \langle{\bf  \theta}^T \rangle = {\bf F}^{-1}
\ee
where
\be
F_{ij} \equiv {1\over 2}{\rm Trace}\left({\bf C}^{-1}{\bf C}_{,i}{\bf C}^{-1}
{\bf C}_{,j}\right)
\ee
is the Fisher information matrix.
{\bf C} is the covariance matrix of the data $\langle {\bf DD}^T\rangle$,
and ${\bf C}_{,i} \equiv \partial {\bf C}/\partial \theta_i$.  The
Fisher matrix is readily obtained from the data covariance matrix (16).
To illustrate this, we calculate the parameter covariance matrix for a 
thin slice, $6^\circ \times 90^\circ$, with a Gaussian selection function 
$\rho_0(r) = \rho_* \exp(-r^2/r_*^2)$.  We take $\rho_* = 0.02 h^3$ Mpc$^{-3}$
and $r_*=450 h^{-1}$ Mpc, broadly comparable to an optical survey 
to a limit $b=19.5$ (cf forthcoming AAT and Sloan galaxy surveys).

We analyse modes from $m=2$ to $m=20$, once again separated by 2 to avoid the 
covariance matrix becoming numerically singular.  The $k$ values are spaced
by 0.0167 $h$ Mpc$^{-1}$, and the modes are analyzed up to 
$k=0.05 h $ Mpc$^{-1}$,
consistent with our previous numerical experiments for unbiased results.
The summations extend to $\ell=60$, and the $k$ integrations
extend to $k=0.165  h$ Mpc$^{-1}$.  The galaxies are weighted with the Feldman 
et al. (1994) optimized weighting $w_s(r) = \left[1+\rho_0(r)
P(k)\right]^{-1}$
and $P(k)$ in the weighting is taken to be 2700 $h^{-3}$ Mpc$^3$.
The expected error on $\beta$ from such a slice is 0.236, and the 
expected fractional error in the power is shown in Fig. 5, for a 
power spectrum assumed to be smooth on a scale of 0.0167  $h$ Mpc$^{-1}$.
Increasing the width of these $k$ bins decreases the error.  Note how the
error increases at the high-$k$ end beyond the maximum wavenumber analysed
(4.0/3.3 times 0.05  $h$ Mpc$^{-1}\ \simeq 0.06$), and at the low-end, 
where the size of the
survey becomes comparable to the wavelength ($2\pi/r_* \simeq 0.014$).
The correlation matrix for the parameters is shown in Table 1.  This
analysis takes only a matter of minutes on a workstation, once the matrices
$\Phi$ and $V$ have been calculated.  These take a few hours,  but are
calculated once only for a given survey.   What is apparent from this
example is that, even for a deep survey with many objects,  $P(k)$ is
detected on scales of the survey $k \sim 2\pi/450 \sim 0.014  h\,$Mpc$^{-1}$,
but not with good accuracy.   $\beta$ estimation is actually not bad 
(error 24\%), but this
could be improved noticeably (to 15\%) if the 3D power spectrum is determined 
independently, from a sky-projected catalogue such as the APM survey.
An application of this method to the Las Campanas survey (Shectman et al. 
1995) is in progress.

\vbox{
\begin{center}
\begin{tabular}{r|r|r|r|r|r}
\\
$\beta$&$P_1$&$P_2$&$P_3$&$P_4$&$P_5$\\
\\

  1.00 &-0.36 &-0.29 &-0.43 &-0.21 & 0.10\\
 -0.36 & 1.00 &-0.21 & 0.21 & 0.03 &-0.04\\
 -0.29 &-0.21 & 1.00 &-0.24 & 0.12 &-0.03\\
 -0.43 & 0.21 &-0.24 & 1.00 &-0.12 &-0.24\\
 -0.21 & 0.03 & 0.12 &-0.12 & 1.00 &-0.58\\
  0.10 &-0.04 &-0.03 &-0.24 &-0.58 & 1.00\\

\\
\end{tabular}
\end{center}
{\bf Table 1.} Correlation matrix for the parameters, in the
order  $\beta$ and the five fractional power spectrum measurements
in order of increasing $k$.\\
}

%
%
\begin{figure}
\begin{center}
\setlength{\unitlength}{1mm}
\begin{picture}(90,70)
\includegraphics{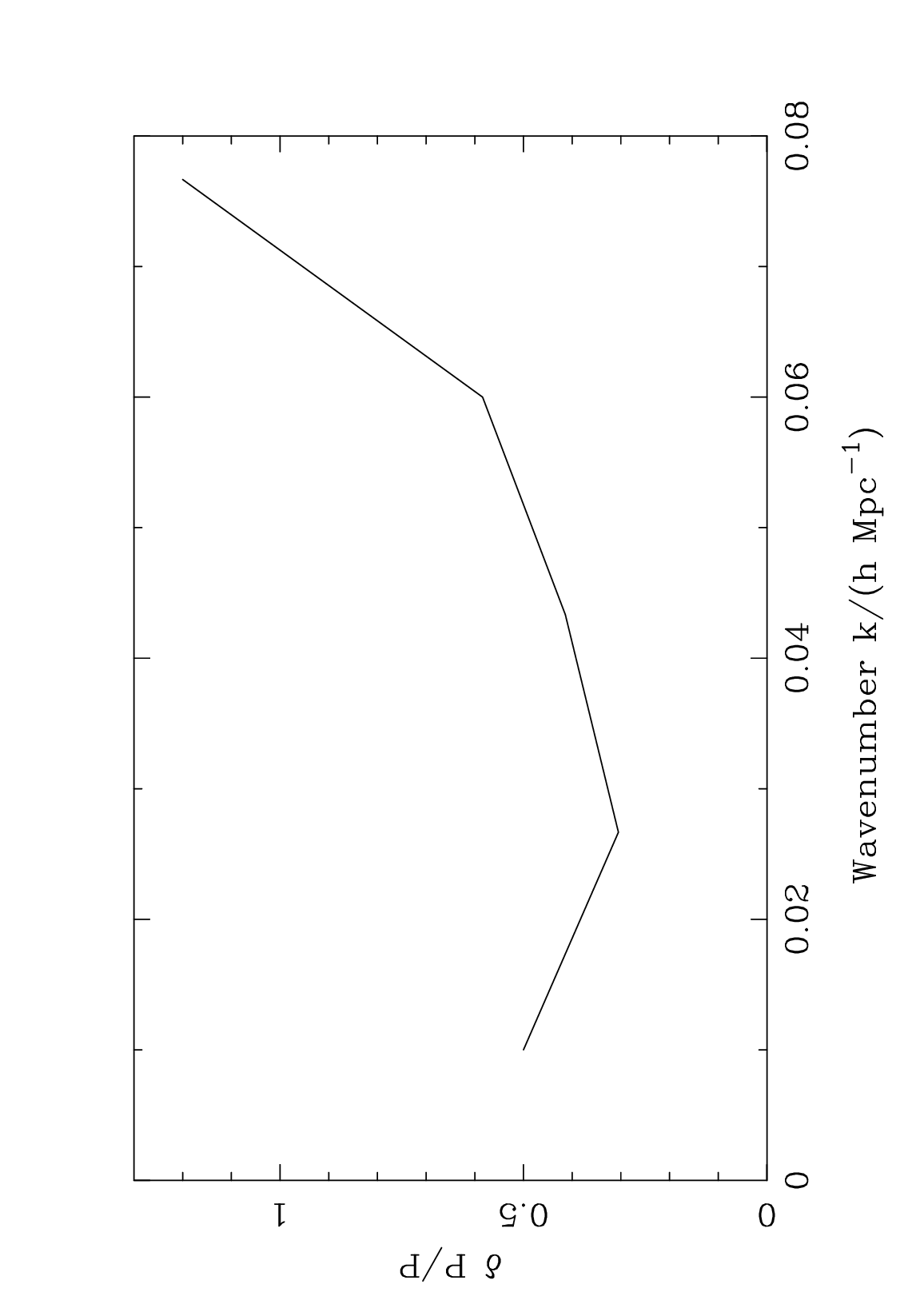} 
\end{picture}
\end{center}
\vspace{2cm}
\caption{Expected fractional error on $P(k)$ from a deep thin slice survey, for
narrow bins in $k$-space.  For details, see text.}
\end{figure}  
%
%

\section{3D surveys: optimizing for power estimation}

We have shown in the previous section how the expected error on 
a parameter may be estimated in advance for a given analysis method and
survey design.  The conclusion that 2D surveys are not particularly good 
for determining large-scale 3D power suggests that genuine 3D surveys may be 
more profitable.  The issues of design and analysis are also relevant in 3D,
and here we consider the problem of optimising the design of a galaxy
redshift survey to measure the power spectrum on some particular 
scale.    The typical decisions to be made are whether to go for a deep
survey over a small area of sky, or a shallower survey over a 
wider area of sky.  We also consider whether it makes sense to
sparse-sample the galaxies, or to observe every one.  This section
is essentially a Fourier analogue of Kaiser's (1986) treatment of
sparse-sampling to estimate the two-point correlation function,
generalized to account for a radial selection function, and with
the proper power error estimate of Feldman, Kaiser \& Peacock (1994)
incorporated.

For simplicity, we make the following assumptions:  we assume
that the effects of redshift-space distortions are small, and
we assume that the observing time for each galaxy is proportional
to the inverse square of its flux.  The former is motivated by
earlier studies (HT95) where the optimal 
weighting was found to be insensitive to the degree of redshift distortion
(see also Hamilton 1996).    The latter assumption is an example;
different constraints, for fibre systems for example,  could be incorporated
if desired.  Our prime constraint is that the total duration of observing
is taken to be fixed.   In considering sparse-sampling, we restrict 
attention to a sparse-sample fraction which is constant for all galaxies
in the parent sample.  Thus, for example, we do not consider a
variable sparse-sampling rate which depends on flux.

It was shown by Feldman et al. (1994), HT95 and Hamilton (1996) that the 
optimal weighting of galaxies in the survey is 
\be
w(\rb) = {1\over 1 + f \bar n(\rb) P(k)}
\ee
where $P(k)$ is the (prior estimate of) the power to be measured,
and $\bar n(\rb)$ is the mean number density of galaxies at position
\rb.  We introduce the possibility of sparse-sampling by multiplying 
this number density by a factor $f$. 

Feldman et al. demonstrated that this weighting gives rise to an error
in the power of $ \sigma_P^2/P^2 = (2\pi)^3/(V_{\bf k}I)$, where
$V_{\bf k}$ is the volume of $k-$space over which the power is
averaged, and
\be
I(f,S,\Omega) =  \Omega \int dr {r^2\over 
\left(1 + {1\over f P\bar n(r)} \right)^2}.
\ee
Here $S$ and $\Omega$ are the flux limit and solid angle of the survey.
Our problem then reduces to maximising $I$ with respect to $f$, $S$
and $\Omega$, subject to the constraint that the total observing time
is fixed.

To do this optimisation, we need the luminosity function $\Phi(L)$,
from which the number density is obtained:
\be
\bar n(r, S) = n_0(X) \equiv \int_{X}^{\infty} dL\, \Phi(L).
\ee
where $X = 4\pi r^2 S$.  If the time to observe an object of 
flux density $S'$ is
$\lambda/(16\pi^2 S'^2)$ for some constant $\lambda$, then the time to 
observe a fraction $f$ of
all objects to a flux limit $S$ in a solid angle $\Omega$ reduces to 
\ba
t(f,S,\Omega) &=& \Omega f\int_0^\infty dr\,\lambda r^6 
\int_{4\pi r^2 S}^{\infty} dL\, {\Phi(L)\over L^2}\nn
&=& {\Omega\lambda f\over 2(4\pi S)^{7/2}} \int_0^\infty dX X^{5/2} n_2(X)\nn
\ea
where $n_2(X) \equiv \int_{X}^{\infty} dL\, \Phi(L)\, L^{-2}$.

The time constraint then simply yields $S \propto (\Omega f)^{2/7}$,
and the error is minimised when
\be
{\Omega^{4/7}\over f^{3/7}} \int dX\,{X^{1/2}\over 
\left(1+{1\over fP n_0(X)}\right)^2}
\ee
is maximised.  $\Omega$ and $f$ may be chosen freely, apart from the 
obvious limits, with the depth of the survey $S$ being dependent on the
choice.  We see immediately that the error is minimised if $\Omega$ is
made as large as possible.  This is a quite general result, 
consistent with the general knowledge that surveys should be wide before
being deep.  If we fix the solid angle of the survey (as large as convenient),
then we can straightforwardly solve for $f$ to optimize the error.
The analysis is readily generalized for observing times which are
proportional to $S'^{-\alpha}r^{-\beta}$ ($\alpha=\beta=2$ might be
appropriate for fixed-width slit spectroscopy).   In this case, one
maximizes
\be
f^{-{1\over 1+2\alpha/3-\beta/3}} \int dX\,{X^{1/2}\over 
\left(1+{1\over fP n_0(X)}\right)^2}.
\ee

Fig. 6 illustrates the effect for a Schechter luminosity function
$\Phi(L)dL = \phi^* (L/L^*)^{-1.3}\,\exp(-L/L^*)dL/L^*$, with 
$\phi^* = 0.013 h^3$ Mpc$^{-3}$.  The optimal sampling occurs at
$fP \simeq 500 h^{-3}$ Mpc$^{3}$, although of course $f$ itself is 
bounded above by unity.   To the left of the minimum, shot noise
becomes dominant, whereas to the right, the extra sampling reduces
the volume observable, so that cosmic variance dominates. 
To estimate $f$, the power at a wavenumber
$k = 0.01$ to $0.1 h$ Mpc$^{-1}$ is about 1000-10000 $h^{-3}$ Mpc$^3$,
depending on the galaxy type and theoretical prejudice (e.g. 
Baugh \& Efstathiou 1993,1994,
Ballinger et al. 1995), which motivates a sparse-sampling strategy of
$f\simeq 0.1$.    
The error rises rapidly if $fP\ls 100$,  so one must take care not
to under-sample.  If the power spectrum on large scales has 
the Zel'dovich form $P \propto k$, a survey to measure very 
large-scale power should be sampled fully.

%
%
\begin{figure}
\begin{center}
\setlength{\unitlength}{1mm}
\begin{picture}(90,70)
\includegraphics{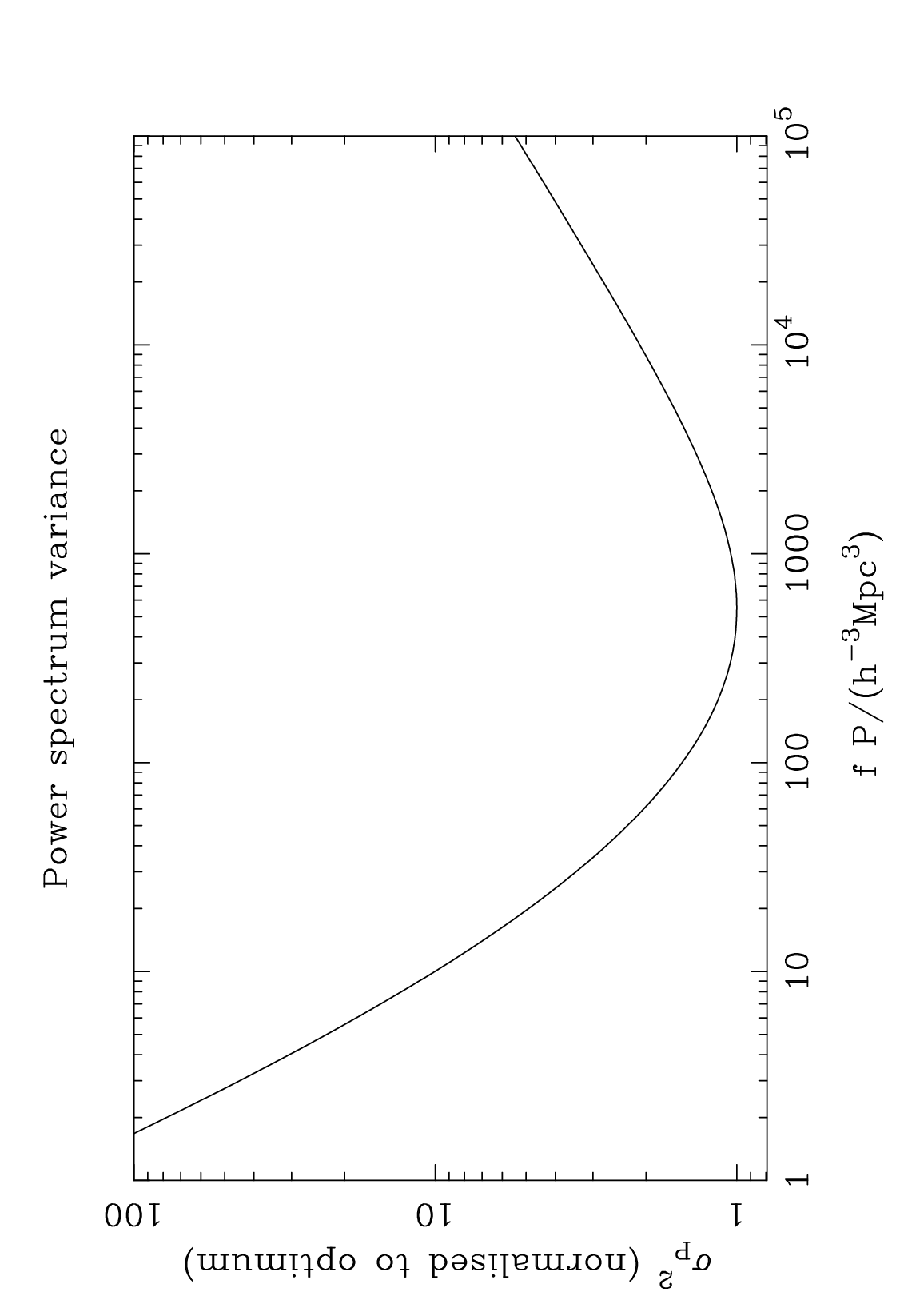} 
\end{picture}
\end{center}
\vspace{2cm}
\caption{The unnormalised variance in the power spectrum for a fixed
solid angle and a Schechter luminosity function, as a function of the
sparse-sample fraction $f$ and the power spectrum to be estimated.  The
details of the luminosity function are given in the text, and the units
of $P$ are $h^{-3}$ Mpc$^3$.}
\end{figure}
%
%

\section{Conclusions}

In this paper we have presented a new method for analysing thin,
near-constant declination slice surveys, using a 2D projection
and expansion in radial and angular functions.  We have also 
considered the optimisation problem of depth and sparse-sampling for
3D surveys.
There are two main advantages of using spherical coordinates 
for analysis.  The first is that the survey 
is usually defined by a fixed areal coverage, and a flux limit 
which leads to a selection function which is purely radial.   
The second advantage is that the effect of redshift distortion 
is radial, so it is straightforward to include it in the analysis.    
By expanding in spherical and angular functions, one can treat 
linear redshift distortions without further approximations, 
and this allows, in particular, 
analysis of long-wavelength modes which do not subtend small
angles on the sky.  By using carefully chosen radial functions
for the analysis, one can ensure that the modes one analyses 
essentially include only 3D modes which are still linear.   

For an all-sky survey, the formalism leads to very simple analysis,
 and this method is clearly the best we have to date.   
What was not clear was whether the method could be adapted 
for surveys of relatively small areal coverage, since the 
mixing of modes of different $\ell$ and $m$ make the expansion
more cumbersome.   This paper shows that, even with thin, 
essentially 2D surveys, one can retain the advantages of the 
spherical expansion without severe additional complexity.   
Our error analysis shows that 3D power 
can be estimated from 2D surveys,  properly including the effects of aliasing,
mode-mode correlations, shot noise and cosmic variance.  
However, the reduction in 
dimensionality means the errors achievable are unlikely to
be very small.   This is in contrast to 3D surveys, where small errors
on the real-space power spectrum can be achieved from the Fourier-Bessel
technique (Ballinger, Heavens \& Taylor 1995).   In 2D, 
$\beta$ can be determined with reasonable accuracy (about 25\%),  but the best 
approach will probably be to use a sky-projected catalogue to 
estimate the 3D power independently, and then to use spherical harmonics
with the slice to measure $\beta$ with higher accuracy (about 15\%).   

We also show in this paper that 3D surveys may be optimised for measuring 
3D power, given a constraint on total observing time, by
choosing as wide an area of sky as possible, and by 
sparse-sampling at a rate which is dependent on the expected power
to be measured.

\section*{APPENDIX}

In this appendix, we calculate the covariance matrix for the 
continuous spherical transform.  
We transform the density field $\delta(\rb) = 
{1\over (2\pi)^3} \int\,d^3\kb\, \delta_\kb\,e^{i\kb.\rb}$
and expand the exponential as a sum of Bessel functions 
\ba
\delta(\rb)& = & {1\over 2\pi^2} \,\int dk d\Omega_\kb 
\delta_\kb\,\times \nn
& & \sum_{\ell m}i^{\ell}\,
j_{\ell}(kr)\,Y_\ell^m(\theta_{\kb},\varphi_{\kb})\,
Y_\ell^{m}(\theta,\varphi)\,k^2\nn
\ea
where $\Omega_\kb = (\theta_\kb,\varphi_\kb)$ is the 
direction of the $\kb$ vector, 
and $(\theta,\varphi)$ is the direction of \rb.  The 
definition of the spherical harmonics used in this 
paper is that found in Binney \& Tremaine (1987):
\ba
Y_{\ell}^m(\theta,\phi) & = & \sqrt{ {2\ell+1\over 4 \pi}
{(\ell-|m|)!\over(\ell+|m|)!} }
P_\ell^{|m|}(\cos\theta)\nn
& & \times \exp(im\phi) \times \left\lbrace {(-1)^m \qquad m\ge
0 \atop 1 \qquad m<0 }\right.
\ea
The spherical expansion is
\be
\delta_{\ell m}(k) = \sqrt{{2\over \pi}}\,\int\,\delta(\rb)\,j_\ell(kr)\,
Y_\ell^m(\theta,\varphi)\,d^3\rb
\ee
which becomes
\be
\delta_{\ell m}(k) = (2\pi)^{-3/2}\,i^\ell\!\! \int \! d\bar k d\Omega_{\bar\kb}
\,\delta_{\bar\kb} \,Y_\ell^m
(\theta_{\bar\kb},\varphi_{\bar\kb})\,\delta^D(k-\bar k)
\ee
where we have used the orthogonality (Binney \& Quinn 1991)
\ba
\int \,d\Omega dr\,j_{\ell'}(k'r) j_\ell(kr)\,r^2\,
Y_{\ell'}^{m'*}(\theta,\varphi)\, 
Y_\ell^m(\theta,\varphi) & = & \nn
{\pi\over 2 kk'}\delta^D(k-k')\delta^K_{\ell\ell'}\delta^K_{mm'}& & 
\ea
where $\delta^D$ and $\delta^K$ are Dirac and Kronecker delta functions
respectively.  The covariance matrix is
\ba
\langle \delta_{\ell m}(k)\delta^*_{\ell' m'}(k')\rangle & = & (2\pi)^{-3} \,
i^{\ell-\ell'}\,\int \,d\bar k d\bar k' d\Omega_{\bar k} d\Omega_{\bar k'}\nn
& & \!\!\!\!\!\!\!\!\!\!\!\!\!\!\!\!\!\!\!\!\!\!\!\!\!\!\!\!\!\!\!\!\!\! 
\!\!\!\!\!\!\!\!\!\!\!\!\!\!\!\!\!\!\!\!\!\!\!\! 
\langle \delta_{\bar\kb}\delta_{{\bar\kb}'}^*\rangle
Y_\ell^{m*}(\theta_{\bar k},\varphi_{\bar k})
Y_{\ell'}^{m'}(\theta_{\bar k'},\varphi_{\bar k'}) \delta^D(k-\bar k)
\delta^D(k'-\bar k').\nn
\ea
Defining the power spectrum by 
\ba
\langle\delta_{\bar\kb}\delta_{{\bar\kb}'}^*\rangle\!\!\!\! & = &  \!\!\!\!
(2\pi)^3 P(\bar k)
\delta^D(\bar\kb-\bar\kb')\nn
\!\!\!\! &   = & \!\!\!\!(2\pi)^3 P(\bar k)\,{\delta^D(\bar k-\bar k')
\over \bar k^2} 
\delta^D(\mu_{\bar\kb}-\mu_{\bar\kb'})
\delta^D(\varphi_{\bar\kb}-\varphi_{\bar\kb'}),\nn
\ea
($\mu \equiv \cos\theta$) the analagous expression of orthogonality for the
Fourier-Bessel modes is
\be
\langle \delta_{\ell m}(k)\delta^*_{\ell' m'}(k')\rangle =  P(k) 
{\delta^D(k-k')\over k^2}\,\delta^K_{\ell\ell'}\delta^K_{mm'}.
\ee
The power is evenly divided between real and imaginary parts, except for
$m=0$ modes, which are real.\\

\noindent{\sl Redshift distortions}\\

In a redshift space map, galaxies are placed at a position
\sb=($s,\theta,\phi$), where the distance coordinate $s$ is the
recession velocity divided by the Hubble constant $H_0$.  In
general this is not the true distance because the galaxy may have
a peculiar velocity \bfv.  The redshift space position is then
related to the real-space position \rb\ by
\begin{equation}
\sb(\rb) = \rb\left[1+{\left(\bfv-\vlg\right)\cdot \rb \over H_0 r^2}\right]
\end{equation}
where $\vlg$ is the peculiar velocity locally.

To expand the spherical expansion to linear order, we first note that 
$\rho(\rb)d^3\rb = \rho(\sb)d^3\sb$, and make a Taylor expansion of the
 resulting integrand to first order in $s-r$
\be
j_\ell(ks)w_s(s) \simeq j_\ell(kr)w_s(r) + (s-r){d\over dr}\left[
j_\ell(kr)w_s(r)\right].
\ee
To obtain an expression for $s-r$, we assume potential 
flow $\bfv=-\nabla\Phi$ (valid for linear, growing-mode 
perturbations), where $\Phi(\rb)$ is the velocity potential.  
The effect of the local group velocity ${\bf v}_0$ is to add an extra term
to the mean of the transform coefficients, and will be treated separately.
Expanding $\Phi$ in terms of $\Phi_{\ell m}(k)$, we find 
\ba
s-r & = & {\bfv.\hat\rb\over H_0}\nn
& = & -{1\over H_0}\sqrt{{2\over \pi}}\sum_{\ell m}\int\,dk 
\Phi_{\ell m}(k) {dj_\ell(kr)\over dr} Y_\ell^{m*}
(\theta,\varphi)\,k^2.\nn
& & 
\ea
The peculiar Poisson equation $\nabla^2\Phi = \beta\delta(\rb)$ 
relates the potential to the galaxy overdensity field, which 
leads to $\Phi_{\ell m}(k)=-\beta \delta_{\ell m}(k)/k^2$.  
 From this we find (choosing units such that $H_0=1$)
\be
s-r = \beta\sqrt{{2\over \pi}}\sum_{\ell m}\int \,dk
\,\delta_{\ell m}(k) {dj_\ell(kr)\over dr} Y_\ell^{m*}(\theta,\varphi).
\ee
This leads to the $V$ matrix terms in the main text.
The effect of the local group velocity is to add the following to the mean 
value
of $D_m(k)$:
\ba
{v_0\over H_0}\sqrt{2\over \pi}\int dr \left(r^2 {d\rho_0\over dr}+2r\rho_0\right)
w_s(r)f(kr)\nn
\int d\Omega\, w_\Omega(\Omega)\exp(-im\varphi) \hat{\bf v}_0\cdot 
\hat{\bf r}
\ea
where $\hat{\bf r}$ and $\hat{\bf v}_0$ are unit vectors.
\\

\noindent{\bf REFERENCES}
\bib \strut

\bib Ballinger W.E., Heavens A.F., Taylor A.N., 1995, MNRAS, 276, L59

\bib Baugh C.M., Efstathiou G.P., 1993, MNRAS, 265, 145

\bib Baugh C.M., Efstathiou G.P., 1994, MNRAS, 267, 323

\bib Binney J., Quinn T., 1991, MNRAS, 249, 678

\bib Binney J., Tremaine S., 1987, Galactic Dynamics, 
Princeton University Press, Princeton

\bib Broadhurst T.J., Ellis R.S., Koo D.C., Szalay A.S., 1990, Nat, 343, 726

\bib Cole S., Fisher K., Weinberg D., 1994, MNRAS, 267, 785

\bib Couchman H.M.P., 1991, ApJ, 368, 23

\bib Fabbri R., Natale V. 1990, ApJ, 363, 3

\bib Feldman H.A., Kaiser N., Peacock J.A., 1994, ApJ, 426, 23

\bib Fisher K.B., Scharf C.A, Lahav O., 1994a, \mn,  266, 219

\bib Fisher K.B., Davis M., Strauss M.A. , Yahil A., Huchra J.P., 
1994b, MNRAS, 267, 927

\bib Fisher K.B., Lahav O., Hoffman Y., Lynden-Bell D., Zaroubi S.,  
1995, MNRAS, 272, 885

\bib Hamilton A., 1992, ApJ, 385, L5

\bib Hamilton A., 1996, MNRAS, in press

\bib Heavens A.F., Taylor, A.N., 1995,  MNRAS, 275, 483


\bib Kaiser N., 1986, MNRAS, 219, 785

\bib Kaiser N., 1987, {\mn}, 227, 1

\bib Kaiser N., Peacock J.A., 1991, ApJ, 379, 482

\bib Landy S.D., Shectman S.A., Lin H., Kirshner R.P., Oemler A.A., 
Tucker D.,  1996, ApJ, 456, L1

\bib Lahav O.,  1993, in Proc. 9th IAP Conference, Cosmic Velocity 
Fields, ed. Bouchet, F., Lachi\' eze-Rey, M., Editions-Fronti\`eres, 
Gif-Sur-Yvette







\bib Scharf C.A., Hoffman Y., Lahav O., Lynden-Bell D., 1992, 
MNRAS, 259, 229

\bib Scharf C.A., Lahav O., 1993, MNRAS, 264, 439

\bib Shectman S.A., Landy S.D., Oemler A., Tucker D.L., Kirshner R.P.,
Lin H., Schechter P.L., 1995, in Wide-Field Spectroscopy and the Distant 
Universe, Proc 35th Herstmonceux Conference, ed. Maddox S.J., 
Aragon-Salamanca A. (Singapore, World Scientific), 98
 
\bib Zaroubi S., Hoffman Y., 1996, ApJ, 462, 25

\end{document}